\documentclass[trackchanges,twocolumn]{aastex701}

\usepackage{amsmath}

\begin{document}

\renewcommand{\[}{\begin{equation}}
\renewcommand{\]}{\end{equation}}
\def\ba{\begin{eqnarray}}
\def\ea{\end{eqnarray}}

\def\p{\partial}\def\i{{\rm i}}
\def\Rb{R_{\rm b}}
\def\rd{\color{red}}

%
\font\gkvecten=cmmib10
\font\gkvecseven=cmmib7
\let\boldgrk=\gkvecten
\let\boldgrksc=\gkvecseven

\def\gkthing#1{{\mathchoice%
	{\hbox{{\boldgrk\char#1}}}
	{\hbox{{\boldgrk\char#1}}}
	{\hbox{{\boldgrksc\char#1}}}
	{\hbox{{\boldgrksc\char#1}}}}}

\def\valpha{\gkthing{11}}
\def\vbeta{\gkthing{12}}
\def\vgamma{\gkthing{13}}
\def\vdelta{\gkthing{14}}
\def\vepsilon{\gkthing{15}}
\def\vzeta{\gkthing{16}}
\def\veta{\gkthing{17}}
\def\vtheta{\gkthing{18}}
\def\viotaeta{\gkthing{19}}
\def\vkappa{\gkthing{20}}
\def\vlambda{\gkthing{21}}
\def\vmu{\gkthing{22}}
\def\vnu{\gkthing{23}}
\def\vxi{\gkthing{24}}
\def\vpi{\gkthing{25}}
\def\vrho{\gkthing{26}}
\def\vsigma{\gkthing{27}}
\def\vtau{\gkthing{28}}
\def\vupsilon{\gkthing{29}}
\def\vphi{\gkthing{30}}
\def\vchi{\gkthing{31}}
\def\vpsi{\gkthing{32}}
\def\vomega{\gkthing{33}}
{\newif\ifnotend
\notendtrue
\def\veclist{ABCDEFGHIJKLMNOPQRSTUVWXYZabcdefghijklmnopqrstuvwxyz.}
\def\top#1#2.{#1}
\def\tail#1#2.{#2.}
\loop\expandafter\xdef\csname v\expandafter\top\veclist\endcsname%
{{\noexpand\bf\expandafter\top\veclist}}
\edef\veclist{\expandafter\tail\veclist}
\if\veclist.\notendfalse\fi\ifnotend\repeat}
{\newif\ifnotend
\notendtrue
\def\callist{ABCDEFGHIJKLMNOPQRSTUVWXYZ.}
\def\top#1#2.{#1}
\def\tail#1#2.{#2.}
\loop\expandafter\xdef\csname c\expandafter\top\callist\endcsname%
{{\noexpand\cal\expandafter\top\callist}}
\edef\callist{\expandafter\tail\callist}
\if\callist.\notendfalse\fi\ifnotend\repeat}
{\newif\ifnotend
\notendtrue
\def\tallist{ABCDEFGHIJKLMNOPQRSTUVWXYZabcdefghijklmnopqrstuvwxyz.}
\def\top#1#2.{#1}
\def\tail#1#2.{#2.}
\loop\expandafter\xdef\csname t\expandafter\top\tallist\endcsname%
{{\noexpand\tilde\expandafter\top\tallist}}
\edef\tallist{\expandafter\tail\tallist}
\if\tallist.\notendfalse\fi\ifnotend\repeat}
\def\rhop{\rho^{(\alpha)}}
\def\Phip{\Phi^{(\alpha)}}\def\Phipp{\Phi^{(\alpha')}}\def\Phipps{\Phi^{(\alpha')*}}
\def\Phib{\Phi^{(\beta)}}\def\Phibp{\Phi^{(\beta')}}
\def\rhopp{\rho^{(\alpha')}}\def\Phips{\Phi^{(\alpha)*}}
\def\ex#1{\left<#1\right>}
\def\bra#1{\langle#1|}
\def\ket#1{|#1\rangle}
\def\la{\langle}
\def\ra{\rangle}
\def\cEx{\cE_\xi }
\def\real{\Re\hbox{{\rm e}}}                   
\def\imag{\Im\hbox{{\rm m}}}                   
\def\d{{\rm d}}
\def\Vc{v_{\rm c}}
\def\cJ{{\cal J}}
\def\cR{{\cal R}}\def\cI{{\cal I}}\def\cK{{\cal K}}
\def\kB{k_{\rm B}}
\def\mnras{MNRAS}
\def\araa{AnnRA\&A}
\def\apj{ApJ}
\def\apjs{ApJS}
\def\apjl{ApJL}
\def\nat{Nat}
\def\aj{AJ}
\def\aap{A\&A}
\def\aaps{A\&AS}
\def\pasp{PASP}
\def\jcap{JCAP}
\def\pasj{PASJ}
\def\pre{Phys. Rev. E}
\def\sovast{SovA}
\def\df{\textsc{df}}
\def\bolth{\mbox{\boldmath$\theta$}}
\def\bolOm{\mbox{\boldmath$\Omega$}}
\def\vOmega{\bolOm}
\def\vDelta{\mbox{\boldmath$\Delta$}}
\def\bolmu{\mbox{\boldmath$\mu$}}
\def\LCDM{$\Lambda$CDM}
\def\LOSI{\mathrm{LOSI}}
\def\like{\mathcal{L}}
\def\sel{\mathrm{S}}
\def\Gyr{\,\mathrm{Gyr}}
\def\Myr{\,\mathrm{Myr}}
\def\kpc{\,\mathrm{kpc}}
\def\mix{\mathrm{mix}}
\def\res{\mathrm{res}}
\def\cut{\mathrm{cut}}
\def\const{\mathrm{const}}
\def\kms{\,\mathrm{km\,s}^{-1}}
\def\km{\,\mathrm{km}}
\def\masyr{\,\mathrm{mas\,yr}^{-1}}
\def\mas{\,\mathrm{mas}}
\def\Gevdens{\,\mathrm{GeV\,cm}^{-3}}
\def\msun{\,{\rm M}_\odot}
\def\vsol{\mbox{\boldmath{$v_\odot$}}}
\def\vsfr{\mbox{\boldmath{$v_{\rm SFR}$}}}
\def\vlos{{v_\parallel}}
\def\pc{\,\mathrm{pc}}
\def\e{\mathrm{e}}\def\s{\mathrm{s}}
\def\vnabla{{\bf\nabla}}
\def\Rc{R_\mathrm{c}}\def\RL{R_\mathrm{L}}\def\Rd{R_\mathrm{d}}
\def\fracj#1#2{{\textstyle{#1\over#2}}}
\def\HI{H\textsc{I}}
\def\codename{\textsc{tm}}
\def\rs{r_{\rm s}}
\def\zs{z_{\rm s}}
\def\tolJ{\textsc{tol}_J}
\def\rms{\textsc{rms}}
\def\vthetaT{\vtheta^{\rm T}}\def\vJT{\vJ^{\rm T}}
\def\thetaT{\theta^{\rm T}}\def\JT{J^{\rm T}}\def\PhiT{\Phi^{\rm T}}
\def\vxT{\vx^{\rm T}}\def\vvT{\vv^{\rm T}}
\def\rT{r^{\rm T}}\def\varthetaT{\vartheta^{\rm T}}\def\pT{p^{\rm T}}
\def\TM{{\sc tm}}
\def\sgn{{\rm sgn}}
\def\figref#1{Fig.~\ref{#1}}
\def\sos{{\sc sos}}
\def\tolJu{\kpc^2\Myr^{-1}}
\def\vomegap{\vomega_{\rm p}}
\def\omegap{\omega_{\rm p}}
\def\Rg{R}\def\Omg{\Omega} 
\def\cL{{\cal L}}
\def\eqrf#1{(\ref{#1})}

\title{Velocities of Free Floaters in a Sea of Stars}

\author{Jun Yan Lau}
\affiliation{Tsung-Dao Lee Institute, Shanghai Jiao Tong University, Shanghai, 201210, China}
\email[show]{junlau@sjtu.edu.cn}  

\author{Dong Lai} 
\affiliation{Tsung-Dao Lee Institute, Shanghai Jiao Tong University, Shanghai, 201210, China}
\affiliation{Center for Astrophysics and Planetary Science, Department of Astronomy, Cornell University, Ithaca, NY 14853, USA}
\email[show]{donglai@sjtu.edu.cn}

\begin{abstract}
We investigate the velocity evolution of free-floating planets and
interstellar objects (``free floaters'') through gravitational
scatterings by field stars (with the stellar mass $m$ much larger than
the mass of the floater, $m_p$).  We show that the equilibrium
velocity -- where dynamical friction balances
stochastic acceleration -- is given by $\sigma                      
\sqrt{2\ln(m/m_p)}$ (where $\sigma$ is the velocity disperson of the field
stars), diverging from the standard energy equipartition
scaling. While the timescale to reach this equilibrium is
prohibitively long, we find that slow floaters ($v \lesssim \sigma$)
undergo mass-independent acceleration, doubling their velocities
within a few relaxation times. Consequently, free floaters initially
following the Maxwellian distribution of their parent stars develop
distinctly non-Maxwellian velocity distributions on a relaxation
timescale.  Since the relaxation time of the Galactic disk is longer than the
age, our results suggest that the kinematics of low-mass free
floaters in the disk may preserve signatures of their parent stars and 
ejection history.
\end{abstract}


\section{Introduction}

Free floating planets (FFPs) are planet-mass objects which are not bound to any host star. They can be studied via direct imaging and microlensing surveys. In recent decades, hundreds of FFP candidates have been detected \citep{ZapateroOsorio2000,Sumi2011,Mroz2017,Mroz2019b,Mroz2020, Miret-Roig2022,Gould2022,Sumi2023}, with masses ranging from sub-Earth mass to tens of Jupiter masses.

Interstellar objects (ISOs) are the asteroid/meteor equivalents of FFPs\citep{JewittSeligman2023}. Thus far only three ISOs have been observed to pass through our Solar System on a hyperbolic trajectory: `Oumuamua \citep{Oumuamua2019}, Borisov  \citep{Borisov2020} and ATLAS \citep{Atlas2025}.

The velocities of ISOs came into the spotlight when ATLAS was observed: it passed through our solar system at $58\kms$, much faster than the local standard-of-rest velocity dispersion $\sigma \approx 20-30 \kms$ \citep{Seabroke2007,Sun2025}, on a trajectory from the Sagittarius constellation, not far from the Galactic centre \citep{Seligman2025}.  Recently, \cite{Subo2026} observed a gravitational microlensing event from both ground and space-based telescopes, and found a Saturn mass FFP with tranverse speed of about $100 \kms$, approximately twice that of the stellar velocity dispersion in the vicinity of the FFP.

These observations raise the question as to whether ISOs and FFPs, henceforth free floaters, share the same velocities as the stars they originate from. 

One factor in determining the velocities of these free floaters is how they are ejected from their parent stars through mechanisms that are connected to the formation processes of the protoplanetary disks and planetary systems around the parent stars. 

The dynamical ejection of free floaters is perhaps the most natural mechanism of producing free floaters, with ejection occuring through planet scatterings\citep{VeraRaymond2012,Ma2016,LiLai2020, Bhaskar2025,HaddenWu2025,HuangLai2025,GuoIda2026}, stellar fly-bys in star clusters \citep{MalmbergDaviesHegggie2011,YuLai2024}, as well as star-planet interactions in circumbinary systems \citep{Sutherland2016,Coleman2024}. A non-exhaustive list of other forms of ejections include post main-sequence escape (owing to stellar mass loss) \citep{Veras2011}, and for massive ``planets'', core collapse akin to brown dwarf formation \citep{PadoanNorlund2002,HennebelleChabriers2008}, ejection of a stellar embryo from a nursery (star abortion) \citep{ReipurthClarke2001} and photo-erosion of a stellar embryo by a OB-type star \citep{WhitworthZinnecker2004}.

While a comprehensive survey of ejection velocities from various mechanisms has not been made, for planet-planet scatterings, \cite{HuangLai2025} used numerical experiments to derive the distribution of ejection velocities for a range of planetary system parameters. For two or three-planet systems on initially circular orbits that experience dynamical instability, the ejection velocity distribution peaks at \begin{equation*}\begin{aligned}
    v_c &\simeq \Bigg({G m_p \over 0.12 a_p}\Bigg)^{1/2}{m_p \over m_p + m_i}\\& = (2.7 \kms) \Bigg({m_p \over m_J}\Bigg)^{1/2}\Bigg({a_p \over 1 \mathrm{AU}}\Bigg)^{-1/2}{m_p \over m_p + m_i}
\end{aligned}\end{equation*}
where $m_p$ is the largest planet mass in the system, $a_p$ is its semi-major axis, and $m_i$ the mass of the ejected planet/object. For two-planet systems, the distribution extends to $\sim 2 v_c$, and for three-planet systems the distribution is broader, extending to $\sim 5 v_c$. For ejections of circumbinary planets, \cite{HuangLai2025} find a similar characteristic ejection velocity
\begin{equation*}\begin{aligned}
    v_c &\simeq \Bigg({G m_0 \over 0.12 a_p}\Bigg)^{1/2}{m_0 \over m_0 + m_1} \\&= (87 \kms) \Bigg({m_0 \over m_\odot}\Bigg)^{1/2}\Bigg({a_p \over 1 \mathrm{AU}}\Bigg)^{-1/2}{m_0 \over m_0 + m_1} 
\end{aligned}\end{equation*}
where $m_0$ and $m_1$ are the primary and secondary stellar masses of the binary. For planet ejections triggered by stellar flyby, \cite{YuLai2024} find that the ejection velocity peaks at \begin{equation*}
    v_c \simeq \Bigg({G m_* \over a_p}\Bigg)^{1/2} = (30 \kms) \Bigg({m_* \over m_\odot}\Bigg)^{1/2}\Bigg({a_p \over 1 \mathrm{AU}}\Bigg)^{-1/2}.
\end{equation*}
Thus, with the exception of circumbinary planets or rare stellar flybys, most floaters produced by planet scatterings would have small ejection velocities (a few $\kms$). In the following, we will neglect the ejection velocity. 

After ejection, the trajectories of both free floaters and their parent stars evolve in tandem under the influence of the Galactic potential and large-scale substructures like spiral arms and the bar\citep{Forbes2025,Bonaca2025,FieldofStreams}. In thispicture, the present day free floater velocity distribution should be approximately equal to the velocity distribution of their parent stars, since both stars and free floaters are treated equally by large-scale potential disturbances in the Galaxy.

This picture is incomplete, however. Free floaters may experience gravitational scatterings by the background stars, such that over time, their velocities are modified. In this paper, we quantify the effect of these scatterings on the evolution of velocity distributions of free floaters with a simple model: that of a free floater drifting in an infinite sea of stars. 

A naive ``thermodynamics argument'' might suggest that given enough time, the free floater (mass $m_p$) will attain an ``equipartition'' velocity $v_{\mathrm{eq}}$ given by $m_p v_{\mathrm{eq}}^2 \sim m \sigma^2$, where $m$ and $\sigma$ are the background stellar mass and velocity dispersion. We show that while this is true for $m_p \gtrsim m$, it is incorrect for free floaters with $m_p \ll m$.

Section 2 reviews the dynamics of gravitational scattering. In Section 3 we compute the ``terminal'' equilibrium velocity of free floaters in an infinite sea of stars, and determine the time evolution of the velocity. Section 4 presents the evolution of the velocity distribution of free floaters as they are produced from an early time to the present day. We conclude in Section 5.

\section{Gravitational Scattering}

We follow \cite{Chandra1949}'s prescription for gravitational scattering between a particle (mass $m_p$) and an infinite, spatially homogeneous medium of stars (mass $m$), which we will refer to as field/background stars. The background stars have a number density $n$, and the velocity distribution is Maxwellian, $f_0 \propto \exp(-{\vv^2 \over 2\sigma^2})$ with a velocity dispersion $\sigma$. This particle will undergo one-on-one hyperbolic scattering in independent, spaced out gravitational encounters with the field stars, never encountering the same star twice. Each encounter modifies the particle velocity, $\vv$, by a factor $\delta \vv$. Neglecting strong encounters with, $|\delta \vv|/v \sim 1$, which comprise the minority of collisions, the net effect of scatterings can be understood by summing up the effects of the many weak encounters ($|\delta \vv|/v \ll 1$) that the particle experiences. 

Chandrasekhar shows that this summation produces three unique diffusion coefficients describing the average changes per unit time due to a large number of encounters\citep[see][Chapter 8.1]{GDII}. The first corresponds to the dynamical friction, a drag force on $m_p$ that acts in the opposite direction to the velocity,
\[
D(\Delta \vv_\parallel) = -{4\pi G^2 m n (m + m_p)\ln \Lambda \over \sigma^2}G(\tilde{v}).
\]
The other two correspond to the stochastic forcing, an acceleration term akin to Brownian motion in the direction parallel to the velocity of the particle,
\[
D(\Delta \vv_\parallel^2) = {4 \sqrt{2}\pi G^2 n m^2 \ln\Lambda \over \sigma}{G(\tilde{v})\over \tilde{v}},
\]
and perpendicular to it,
\[
D(\Delta \vv_\perp^2) = {4\sqrt{2}\pi G^2 n m^2 \ln \Lambda \over \sigma}\Bigg[{\mathrm{erf}(\tilde{v}) - G(\tilde{v}) \over \tilde{v}}\Bigg].
\]
Here $\ln\Lambda$ is the Coulomb logarithm, \[\tilde{v} = v/(\sqrt{2}\sigma)\] is the dimensionless velocity, and the function $G(\tilde{v})$ is
\[
G(\tilde{v}) = {1 \over 2 \tilde{v}^2}\left[\mathrm{erf}(\tilde{v}) - {2 \tilde{v} \over \sqrt{\pi}}\exp\left({-\tilde{v}^2}\right)\right].
\]

The rate of change of the kinetic energy $K = m_p \vv^2/2$ of the particle, averaged over many weak encounters with the field stars, is given by 
\[\label{eq:K}
{\p K\over \p t} = m_p \left[v D(\Delta \vv_\parallel) + {1 \over 2}\Big(D(\Delta \vv_\perp^2) + D(\Delta \vv_\parallel^2)\Big)\right].
\]
To proceed, we introduce the dimensionless time $\tilde{t} = t/t_r$, where $t_r$ is the relaxation time \footnote{Note that $t_r$ is about a factor of $3$ smaller than the nominally defined relaxation time, Eq.~(7.106) of \cite{GDII}.},
\[\begin{aligned}\label{eq:t_r}
t_r &\equiv {1 \over 4\sqrt{2\pi}\ln \Lambda}{\sigma^3 \over n(\mathrm{G} m)^2}\\
&\simeq (0.3 \Gyr) \left({\ln \Lambda \over 15}\right)^{-1} \left({n \over 1 \pc^{-3}}\right)^{-1} \\& \times \left({\sigma \over 1 \kms}\right)^3 \left({m \over \msun}\right)^{-2}.
\end{aligned}\]

Equation \eqref{eq:K} then becomes
\[\label{eq:DE}
{\p \tilde{v}^2 \over \p \tilde{t}} = {m_p \over m}\Bigg[-{\sqrt{\pi} \over 2\tilde{v}}\mathrm{erf}(\tilde{v}) + \left(1 + {m \over m_p}\right)\exp\left({-\tilde{v}^2}\right)\Bigg].
\]

\section{Equilibrium Velocity}

The particle reaches the ``equilibrium'' velocity $v_{\mathrm{eq}}$ when the energy loss due to dynamical friction balances the energy gain due to (stochastic) kicks from the stars. Setting ${\p \tilde{v} / \p t} = 0$ in Equation \eqref{eq:DE}, we find
\[ \label{eq:zerocdn}
{\sqrt{\pi} \over 2 \tilde{v}_\mathrm{eq}}\mathrm{erf}(\tilde{v}_\mathrm{eq}) \exp\left({\tilde{v}_\mathrm{eq}^2}\right) = 1 + {m \over m_p}.
\]

For free floaters with $m_p \ll m$, this gives
\[
\tilde{v}_{\mathrm{eq}} \approx \sqrt{\ln(m/m_p)},
\]
or the equilibrium velocity $v_\mathrm{eq}$ is
\[
v_\mathrm{eq} \approx \sqrt{2}\sigma \sqrt{\ln(m/m_p)}. 
\]
In contrast, for $m_p \gtrsim 3 m/2$, Eq.~ \eqref{eq:zerocdn} yields
\[
\tilde{v}_\mathrm{eq}^2 \simeq {3 \over 2}{m \over m_p}\left(1 - {3 m \over 5 m_p}\right)
\]
or
\[
{1 \over 2} m_p v_\mathrm{eq}^2 \simeq {3 \over 2}{m \sigma^2}\left(1 - {3 m \over 5 m_p}\right).
\]
Thus, equipartition of kinetic energy applies only when the mass of the particle is larger than the mass of the field star, but not when $m_p \ll m$.

Now that we have obtained the equilibrium velocity of the particle, we can examine how quickly this is achieved. When $m_p \ll m$ and $\tilde{v} < \tilde{v}_\mathrm{eq}$, equation \eqref{eq:DE} simplifies to
\[\begin{aligned}\label{eq:dX^2pre}
{\p \tilde{v}^2 \over \p \tilde{t}}
&\simeq\exp\left({-\tilde{v}^2}\right).\\
\end{aligned}\]
The solution is
\[\label{eq:X_f}
\exp\left(\tilde{v}^2\right) \simeq  \exp\left({\tilde{v}_0^2}\right) + \tilde{t},
\]
where $\tilde{v}_0$ is the initial speed of the particle at time $t = 0$. Thus, a free floater with initial speed $\sigma$ will achieve $2\sigma$ at $\tilde{t}_{2\sigma} = (\e^{4/2} - \e^{1/2}) \approx 5.74$
and will reach $3\sigma$ at $\tilde{t}_{3\sigma} = (e^{9/2} - e^{1/2}) \approx 88.3$.

Note that Eq.~ \eqref{eq:dX^2pre} breaks down when $\tilde{v}$ approaches $\tilde{v}_\mathrm{eq}$. The characteristic timescale to reach the equilibrium velocity is given by
\[
    \tilde{t}_\mathrm{eq} \approx \exp\left(\tilde{v}_\mathrm{eq}^2\right) \approx {m \over m_p}.
\] 
To achieve the equilibrium velocity, a FFP would require $10^3-10^5$ relaxation times. 

\begin{figure}
    \centering
    \includegraphics[width=1\linewidth]{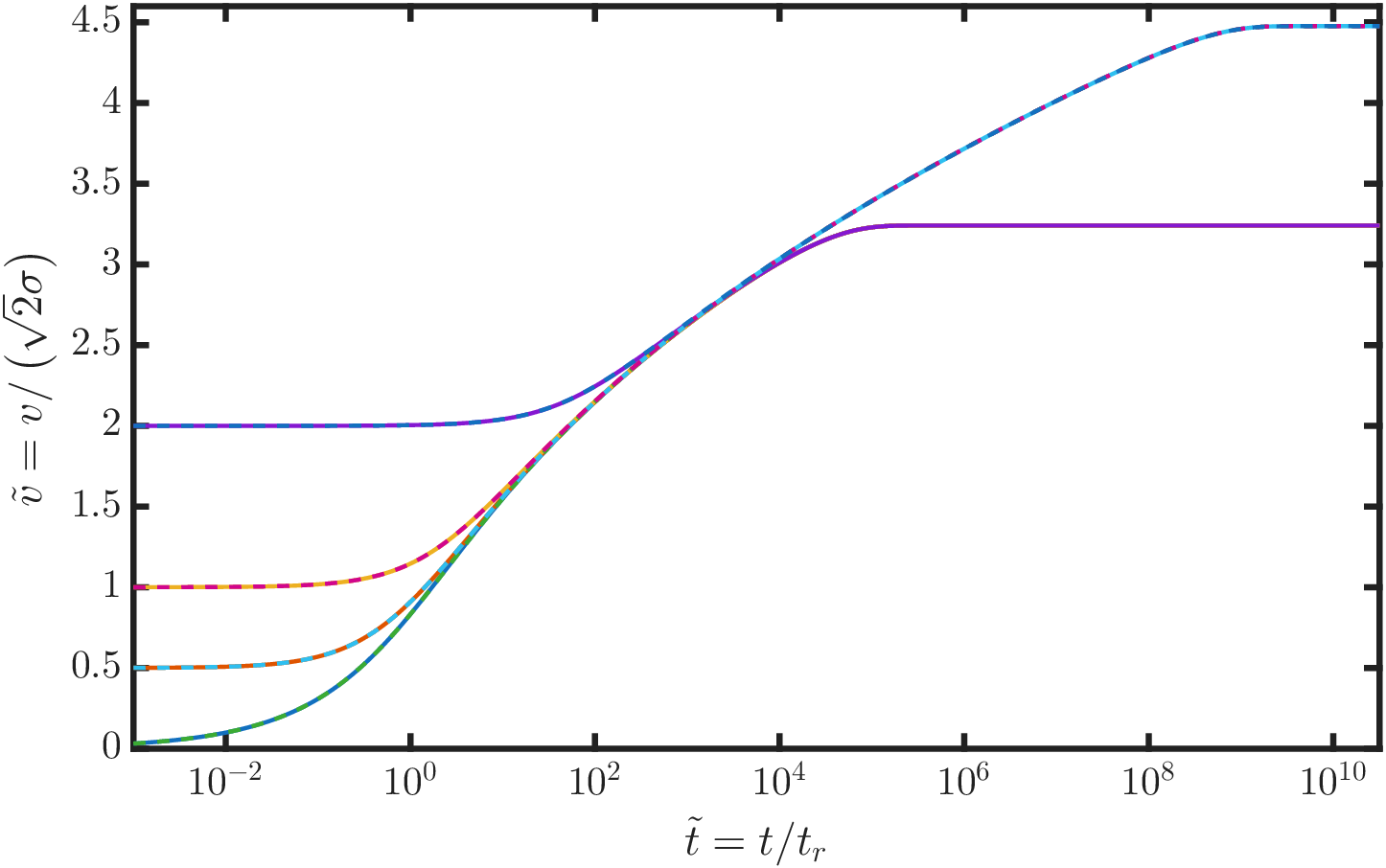}
    \caption{The time evolution of the velocity (in units of $\sqrt{2}\sigma$) of a free floater due to gravitational scatterings by background stars. The time is in units of the relaxation time (see Eq.~\ref{eq:t_r}). The different colored curves correspond to different initial velocities $(\tilde{v}_0 = 0, 0.5, 1,2)$. The solid lines are for $m_p/m = 10^{-4}$, and the dashed lines for $m_p/m = 10^{-8}$. Note that the dashed lines overlap the solid lines before they diverge at $\tilde{t} \sim 10^4$.}
    \label{fig:X_fmin}
\end{figure}

In Figure \ref{fig:X_fmin}, we present the time-evolution of $\tilde{v}$ with respect to $\tilde{t}$ for free floaters with different initial velocities and different masses.

Since $\tilde{t}_\mathrm{eq} \gg 1$ for $m_p / m \ll 1$, we will employ equation \eqref{eq:X_f} in the next section to study how gravitational scatterings influence the velocity distribution of free floaters.

\section{Velocity Distribution Function of Free Floaters}

Gravitational scatterings can modify the distribution function of free floaters in a non-trivial way. When applied to objects that are larger than the field stars, gravitational scatterings produce equilibrium distribution functions that are Maxwellian with the velocity dispersion governed by the equipartition theorem \citep{GDII}. This occurs because these large objects exchange energies with each other via their mutual scattering more efficiently than they do with the field stars, resulting in their rapid thermalisation.

When gravitational scatterings are applied to the free floaters with $m_p \ll m$, this picture is flipped on its head. Scatterings between the free floaters are many orders of magnitude weaker than scatterings between the free floaters and the field stars. Free floaters will never thermalise: they are slaved to the field stars, thus the kinetic energy evolution equation (Eq.~\ref{eq:DE}) actually governs the kinetic energy of each free floater, which compels their distribution function to depart from the Maxwellian.

\subsection{Evolution of Velocity Distribution}

We may compute the distribution function evolution equation by considering the map between a particle's initial velocity $\tilde{v}_0 = \tilde{v}(t=0)$ and its velocity $\tilde{v}$ at the dimensionless time $\tilde{t} = t/t_r$, given by Equation \eqref{eq:X_f}. Assuming that the free floaters have the initial velocity distribution $f_0(\tilde{v}_0)$ given by that of the background stars, we can write the density of the free floaters as
\[\begin{aligned}
n_{\mathrm{float}} &= 4\pi\int_0^{\infty} f_0(\tilde{v}_0) \tilde{v}_0^2 ~\d \tilde{v}_0\\
&= 4\pi \int_{\tilde{v}_{\mathrm{min}}(\tilde{t})}^\infty f_0\Bigl(\tilde{v}_0(\tilde{v},\tilde{t})\Bigr) {1 \over 3} {\d \tilde{v}_0^3 \over \d \tilde{v}} ~\d \tilde{v},\\
\end{aligned}\]
where 
\[\label{eq:vmin}
\tilde{v}_{\mathrm{min}}(\tilde{t}) \equiv \sqrt{\ln(1 + \tilde{t})}
\]
is the minimum velocity at time $\tilde{t}$. We thus identify the (1D) velocity distribution function at time $\tilde{t}$ as \[\label{eq:f_r}F(\tilde{v},t) = f_0(\tilde{v},\tilde{t}) {1 \over 3} {\d \tilde{v}_0^3 \over \d \tilde{v}}.\]

Using $f_0 = \exp(-\tilde{v}_0^2)$ and Equation \eqref{eq:X_f} for the $\tilde{v}_0$---$\tilde{v}$ mapping, we find
\[\begin{aligned}\label{eq:BoseEinstein}
    F(\tilde{v},\tilde{t})
    &= {\tv \exp(-\tv^2) \over (\exp(\tilde{v}^2) - \tilde{t})^2} \left[\ln(\exp(\tv^2) - \tt)\right]^{1/2}.
\end{aligned}
\]
This expression is valid for $\exp(\tv^2)-1\ge \tt$, or $\tv\ge\tv_{\rm min}(\tt)$.
It is easy to check that $F(\tv,\tt=0)=\tv^2\exp(-\tv^2)$, as expected.

\begin{figure}
    \centering
    \includegraphics[width=\linewidth]{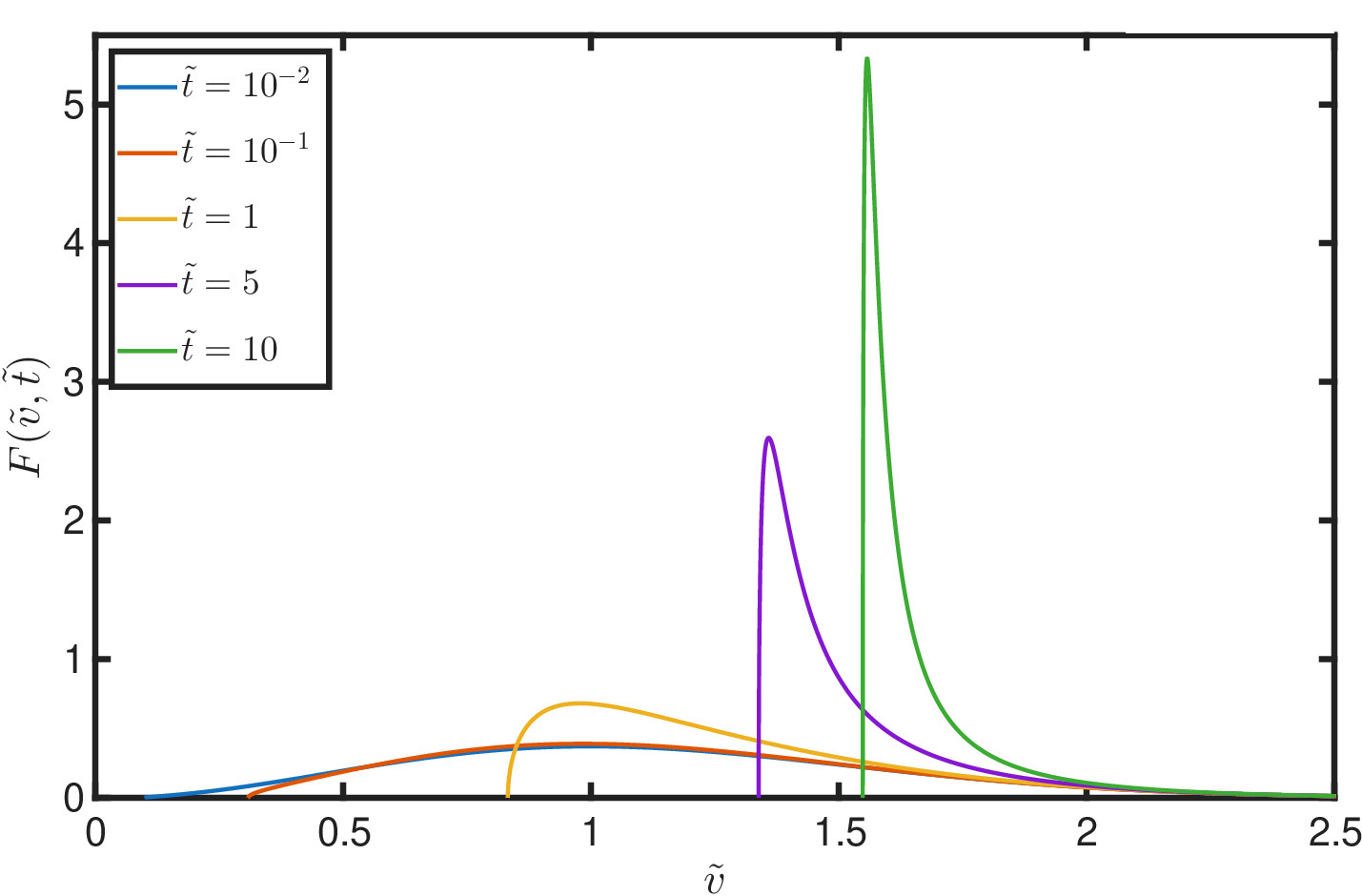}
    \caption{The 1D velocity distribution of $F(\tilde{v},\tilde{t})$ at different times assuming that at $\tilde{t} = 0$ an initial population is born with $f_0 \propto \exp(-\tilde{v}_0^2)$.}
    \label{fig:BE}
\end{figure}

Figure \ref{fig:BE} shows the distribution function $F$ at $\tilde{t} = 10^{-2},10^{-1},1, 5, 10$ after an initial population of free floaters is created with velocity dispersion $\sigma$. 

\subsection{``Observed'' Distribution function}

\def\be{\begin{equation}}
\def\ee{\end{equation}}
\def\ba{\begin{eqnarray}}
\def\ea{\end{eqnarray}}

The present-day (at the Galactic time $T=\tT t_{r}$) ``observed'' free
floaters were produced at earlier times $t_b\in (0,T)$. To determine
their velocity distribution requires that we prescribe a rate at which
free floaters are produced, and their initial velocity distributions
at birth.

For simplicity, here we assume that free floaters are produced at a
constant rate throughout the lifetime of the Galaxy, and with the same
initial velocity distribution. Then the ``observed'' velocity distribution of
all free floaters at time $\tT$ is simply
\be
\bar F(\tv,\tT)={1\over \tT}\int_0^\tT   F(\tv,\tt){\rm d}\tt.
\ee
Using Eq.~\eqref{eq:BoseEinstein}, we have
\be\label{eq:Ft}
\bar F(\tv,\tT)={\tv\exp(\tv^2)\over \tT}\int_U^V {(\ln x)^{1/2}\over x^2}\Theta(x-1)
\,{\rm d} x,
\ee
where $\Theta$ is the step function, and we have defined 
\be
V\equiv \exp(\tv^2),\quad
U\equiv \exp(\tu^2)\equiv \exp(\tv^2)-\tT.
\ee
Note that the integral in Eq.~\eqref{eq:Ft} is
\be
\int {(\ln x)^{1/2}\over x^2}\,{\rm d} x={\sqrt{\pi}\over 2}{\rm erf}(\sqrt{\ln x})
-{\sqrt{\ln x}\over x}.
\ee
We then find
\be
\bar F(\tv,\tT)={\tv^2\over\tT}\left[{\sqrt{\pi}\over 2\tv}\exp(\tv^2)\,
  {\rm erf}(\tv)-1\right]
\ee
if $\tv < \tv_{\rm min}(\tT)$, 
\be\begin{aligned}
\bar F(\tv,\tT)&={\tv^2\over\tT}\bigg[{\tu\over\tv}\exp(\tv^2-\tu^2)-1
  +{\sqrt{\pi}\over 2\tv}\exp(\tv^2)\,\\& \times 
  \Bigl( {\rm erf}(\tv)-{\rm erf}(\tu)\Bigr)\bigg]
\end{aligned}\ee
if $\tv > \tv_{\rm min}(\tT)$, and \[\tv_{\rm min}(\tT) = \sqrt{\ln\left(1 + \tT\right)}\] is given by Eq.~\eqref{eq:vmin}.

\begin{figure}
\centering
\includegraphics[width=1\linewidth]{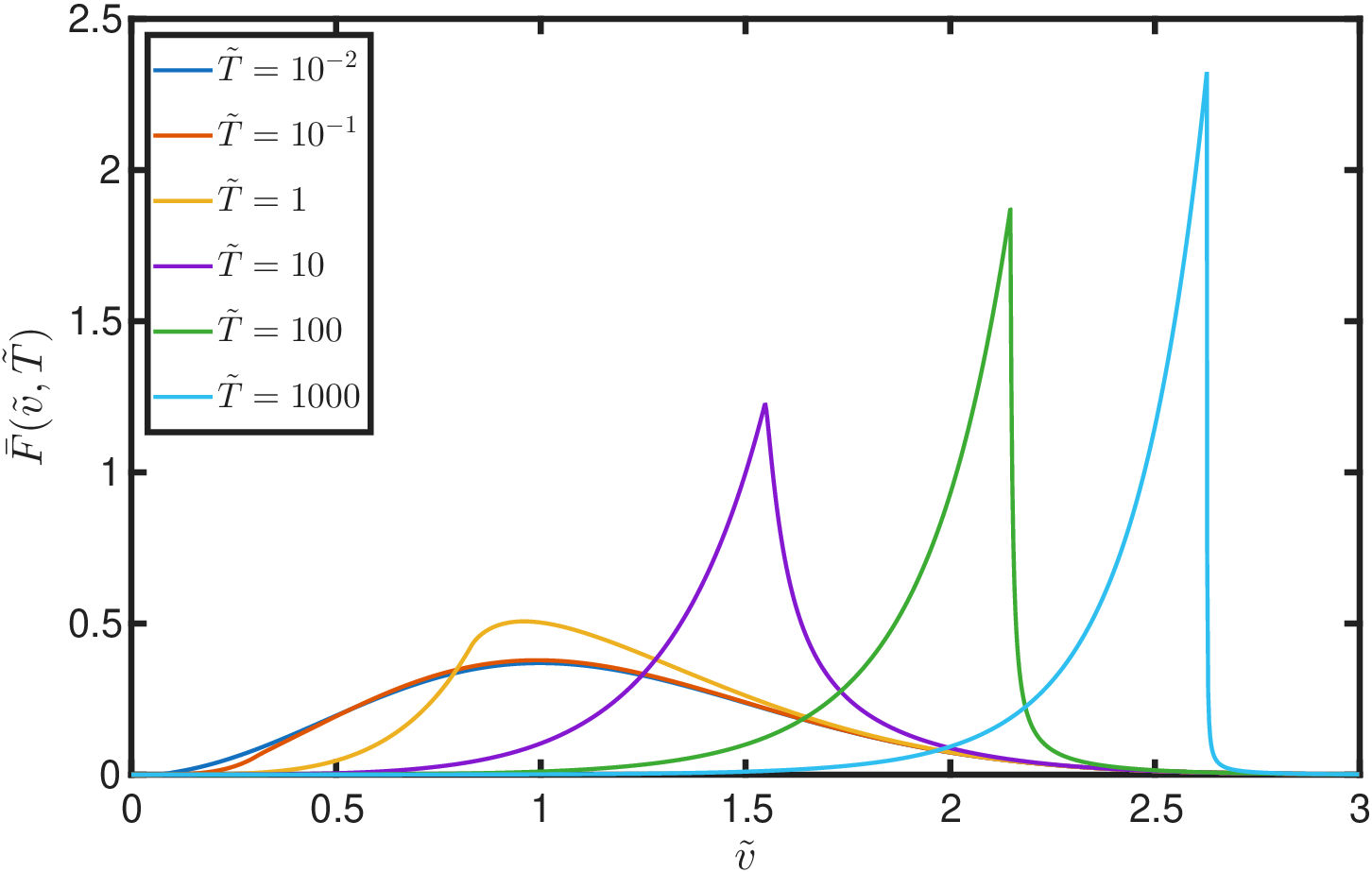}
\caption{The velocity distribution of the ``observed'' free floaters $\bar{F}(\tilde{v},\tilde{T})$ at several different time $\tT$ (as labeled), assuming that the floaters were produced during $\tt\in (0,\tT)$
  at a constant rate with the same initial distribution at birth. The velocity 
  is in units of $\sqrt{2}\sigma$, and time in units of $t_r$ (see Eq.~\ref{eq:t_r}).}
    \label{fig:f_rX}
\end{figure}

In Figure \ref{fig:f_rX}, we show the ``observed'' velocity
distribution $\bar F(\tv,\tT)$ for $\tT =
10^{-3},\,10^{-2},\,10^{-1},\, 1, 10,\,100$.
From $\tT = 1$ onwards, the peaks of the distribution
functions are at $\tilde{v}_{\mathrm{min}}(\tT)$. $\tilde{v}_{\mathrm{min}}(\tT)$ it can be considered the characteristic velocity of free floaters.

\section{Summary and Discussion}

In this paper, we have studied how the velocities and the velocity
distribution function of free floaters (free floating planets and
interstellar objects) evolve due to gravitational scatterings by field
stars, which are much more massive than the floaters ($m\gg m_p$).
Our key results are:
\begin{itemize}
\item  The ``terminal'' equlibirum velocity $v_{\rm eq}$ of the
floater, attained when the dynamical friction balances the
acceleration associated with the stochastic gravitational scatterings
(``Brownian motion'') of the field stars, is given by $\sqrt{2\ln(m/m_p)}$
times the velocity dispersion $\sigma$ of the field stars. This is in contrast to the
$m_p\gtrsim m$ case, where the equilibrium velocity corresponds to
energy equiparition ($v_{\rm eq}^2\simeq 3m\sigma^2/m_p$).
\item The time needed to achieve the equilibrium velocity is approximately equal to $m/m_p$ times
the relaxation time of the field stars (Eq.~6), meaning that very
light free floaters ($m_p\ll m$) would never attain the equilibrium
velocity. Nevertheless, during the initial acceleration phase, a slow
floater (with $v\lesssim \sigma$) can double its velocity within a few
relaxation times, and the velocity evolution is independent of the
mass of the floater (see Eq.~\ref{eq:X_f}).
\item For floaters ejected with the same Maxwellian distribution
as the field stars, the velocity distribution becomes distinctly non-Maxwellian
after a relaxation time.
\end{itemize}

Our calculations are based on a simplified model, i.e., we assume a
homogeneous, infinite sea of field stars. In reality, for a finite
stellar system, the free floater may escape the system when its
velocity exceeds the escape velocity. Note that the escape velocity can be
much larger than the stellar velocity dispersion, if the gravity
of the stellar system has a significant contribution from dark matter
(or the central supermassive black holes in case of nuclear star clusters).

The stellar relaxation time in the solar neighborhood (with $n\sim
0.1$~pc$^{-3}$ for $m\sim M_\odot$, and $\sigma \sim 20$~km/s) is
long, so is the relaxation time in the bulk Galactic bulge (with
$n\sim 10^2-10^3$~pc$^{-3}$ and $\sim 100$~km/s) \citep{GDII}.  
Thus we do not expect gravitational scatterings to
significantly influence the velocities of free floaters produced in
these environments. The kinematics of these floaters may preserve
signatures of their parent stars and the ejection mechanisms.
On the other hand, free floaters produced in dense star clusters
(such as globular clusters or nuclear star clusters) can be strongly
affected by scatterings, as these clusters have shorter relaxation
times.

\bibliographystyle{aasjournalv7}



\end{document}